\documentclass[aps,pre,reprint,superscriptaddress,nofootinbib,floatfix]{revtex4-2}
\usepackage{graphicx}
\usepackage{xcolor}
\usepackage{hyperref}
\usepackage[hyphenbreaks]{breakurl}
\usepackage{amsmath,amssymb,amsfonts,bm}
\usepackage{amsthm}
\usepackage[normalem]{ulem}
\usepackage{lineno}

\def\eqref#1{equation~\ref{#1}}

\def\1{\bm{1}}

\DeclareMathAlphabet{\mathsfit}{\encodingdefault}{\sfdefault}{m}{sl}
\SetMathAlphabet{\mathsfit}{bold}{\encodingdefault}{\sfdefault}{bx}{n}

\newcommand{\model}[0]{PLANCK}

\newcommand{\ENCmodel}[0]{PHGNN}

\begin{document}

\title{Optimizing $p$-spin models through hypergraph neural networks and deep reinforcement learning}

\author{Li Zeng}
\thanks{These authors contributed equally to this work.}
\affiliation{Laboratory for Big Data and Decision, College of Systems Engineering, National University of Defense Technology, Changsha 410073, China}

\author{Mutian Shen}
\thanks{These authors contributed equally to this work.}
\affiliation{Department of Physics, Washington University in St. Louis, Campus Box 1105, 1 Brookings Drive, St. Louis, MO 63130, USA}

\author{Tianle Pu}
\affiliation{Laboratory for Big Data and Decision, College of Systems Engineering, National University of Defense Technology, Changsha 410073, China}

\author{Zohar Nussinov}
\affiliation{Department of Physics, Washington University in St. Louis, Campus Box 1105, 1 Brookings Drive, St. Louis, MO 63130, USA}
\affiliation{Rudolf Peierls Centre for Theoretical Physics, University of Oxford, Oxford OX1 3PU, United Kingdom}

\author{Qing Feng}
\affiliation{Laboratory for Big Data and Decision, College of Systems Engineering, National University of Defense Technology, Changsha 410073, China}

\author{Chao Chen}
\affiliation{Laboratory for Big Data and Decision, College of Systems Engineering, National University of Defense Technology, Changsha 410073, China}

\author{Zhong Liu}
\affiliation{Laboratory for Big Data and Decision, College of Systems Engineering, National University of Defense Technology, Changsha 410073, China}

\author{Changjun Fan}
\email{fanchangjun@nudt.edu.cn}
\affiliation{Laboratory for Big Data and Decision, College of Systems Engineering, National University of Defense Technology, Changsha 410073, China}

\begin{abstract}
$p$-spin glasses, characterized by frustrated many-body interactions beyond the conventional pairwise case ($p>2$), are prototypical disordered systems whose ground-state search is NP-hard and computationally prohibitive for large instances. Solving this problem is not only fundamental for understanding high-order disorder, structural glasses, and topological phases, but also central to a wide spectrum of hard combinatorial optimization tasks. Despite decades of progress, there is still a lack of an efficient and scalable solver for generic large-scale $p$-spin models. Here we introduce \model, a physics-inspired deep reinforcement learning framework built on hypergraph neural networks. PLANCK directly optimizes arbitrary high-order interactions, and systematically exploits gauge symmetry throughout both training and inference. Trained exclusively on small synthetic instances, \model~exhibits strong zero-shot generalization to systems orders of magnitude larger, and consistently outperforms state-of-the-art thermal annealing methods across all tested structural topologies and coupling distributions. Moreover, without any modification, \model~achieves near-optimal solutions for a broad class of NP-hard combinatorial problems, including random $k$-XORSAT, hypergraph max-cut, and conventional max-cut. The presented framework provides a physics-inspired algorithmic paradigm that bridges statistical mechanics and reinforcement learning. The symmetry-aware design not only advances the tractable frontiers of high-order disordered systems, but also opens a promising avenue for machine-learning-based solvers to tackle previously intractable combinatorial optimization challenges.
\end{abstract}

\maketitle

\section*{High-order spin glasses and NP-hard optimization}
\begin{figure*}[t]
\centering
\includegraphics[width=0.95\textwidth]{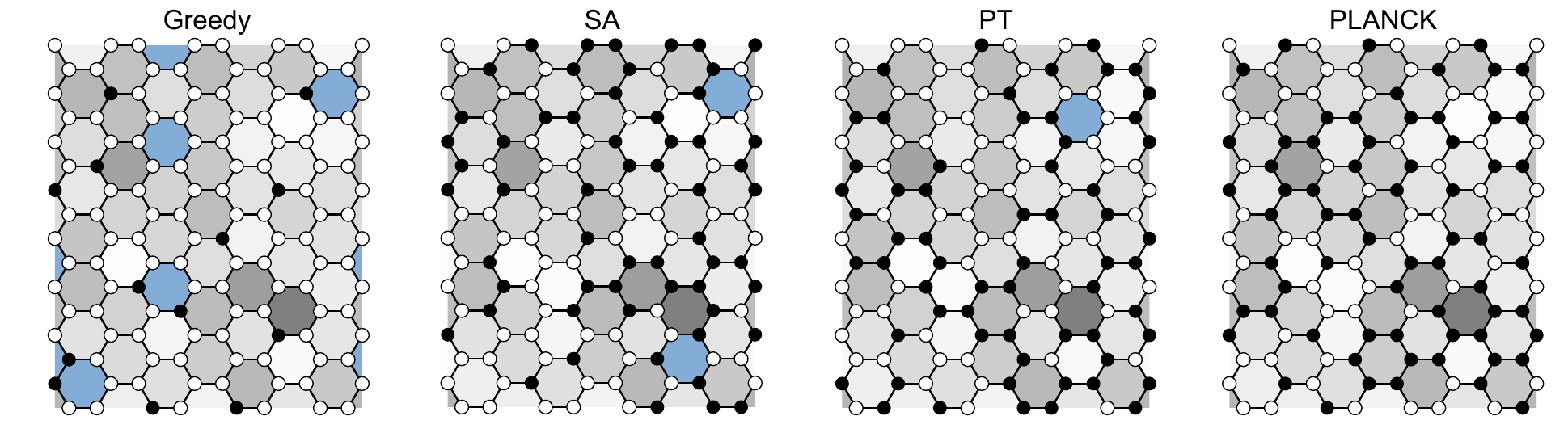}
\caption{Illustrative ground-state optimization on a hexagonal lattice.
Ground‑state search on a $L=8$ hexagonal ($p=6$) lattice with fixed boundary conditions and Gaussian couplings. Each hexagon represents a six‑spin interaction (opacity scales with $|J|$; blue-shaded areas indicate unsatisfied bonds with $-J\Pi\sigma_i > 0$). Snapshots compare the steady-state configurations achieved by Greedy search, Simulated Annealing (SA), Parallel Tempering (PT), and \model. As can be observed, \model~is the only method that reaches the exact ground state (verified by Gurobi, a branch-and-bound based exact solver), whereas other methods stall in higher‑energy local minima.}
\label{fig:case_study_c}
\end{figure*}

Over the past few decades, the spin glass model has become a cornerstone of statistical physics\cite{mezard1987spin,charbonneau2023spin}. A key breakthrough came with Parisi’s replica symmetry breaking (RSB) theory\cite{parisi_numerical_2012}, which provides fundamental insights into the fully connected Sherrington–Kirkpatrick (SK) model\cite{sherrington1975solvable} and extends naturally to more complex disordered systems\cite{gyorgyi2001techniques,ghofraniha2015experimental,pierangeli2017observation}. Gardner further generalized the SK model by introducing $p$-spin interactions\cite{gardner_spin_1985}, where the SK model corresponds to the special case $p=2$ and the random energy model (REM) arises in the limiting case $p\to \infty$\cite{gross_simplest_1984}. Optimizing $p$-spin glass models\cite{thomas_optimizing_2011} with $p>2$ is difficult, particularly for the Edwards–Anderson (EA) model, which describes finite-dimensional lattices with nearest-neighbor interactions\cite{edwards1975theory}. In this regime, mean-field approaches break down, sparking ongoing debates between the RSB framework and the competing droplet model\cite{newman_ground_2021}. At the heart of these challenges lies in minimizing the following $p$-spin Hamiltonian:
\begin{linenomath*}
\begin{equation}\label{eq:hamiltonian}
    \mathcal{H}_p = -\sum\limits_{{i_1} <{i_2} < ... < {i_p}} {{J_{i_1, i_2, \dots, i_p}}} \prod\limits_{k = 1}^p {{\sigma _{{i_k}}}}.
\end{equation}
\end{linenomath*}
In general, this Hamiltonian can be defined on arbitrary graphs. Here we will focus on the most heavily studied Edwards–Anderson (EA) model with Ising spins $\sigma_i=\pm1$ placed on regular two-dimensional lattices—triangular ($p=3$), square ($p=4$), and hexagonal ($p=6$). The coupling $J_{i_1, i_2, \dots, i_p}$ is generally drawn from bimodal or Gaussian distributions (Fig.~\ref{fig:case_study_c} for a hexagonal lattice example).

Optimizing the $p$-spin model with $p>2$ is of fundamental importance for three reasons. First, it is the key to understanding structural glasses\cite{kirkpatrick1987connections}, supercooled liquids\cite{larson2010numerical}, and topological quantum error correction\cite{takada2024ising} governed by high-order disorder. Second, optimizing the native high-order Hamiltonian, rather than a quadratized proxy, eliminates auxiliary variables and retains the original interaction geometry, both of which simplify the problem and improve computational efficiency\cite{bybee2023efficient}. Third, the $p$-spin formulation exactly encodes a vast family of NP-hard combinatorial tasks, from pure-$p$ systems (random $k$-XORSAT\cite{nikhar2024all}, Modern Hopfield model\cite{bovier2001spin}) to mixed-$p$ (hypergraph MaxCut\cite{veldt2022hypergraph}, MAXSAT\cite{molnar2018continuous}) and even conventional pairwise problems (MaxCut\cite{lucas2014ising}, vertex cover\cite{dinur2005hardness}). These characteristics position the $p$-spin model as a universal and physically grounded benchmark for assessing optimization algorithms across diverse regimes.

\begin{figure*}[t]
\centering
\includegraphics[width=0.95\textwidth]{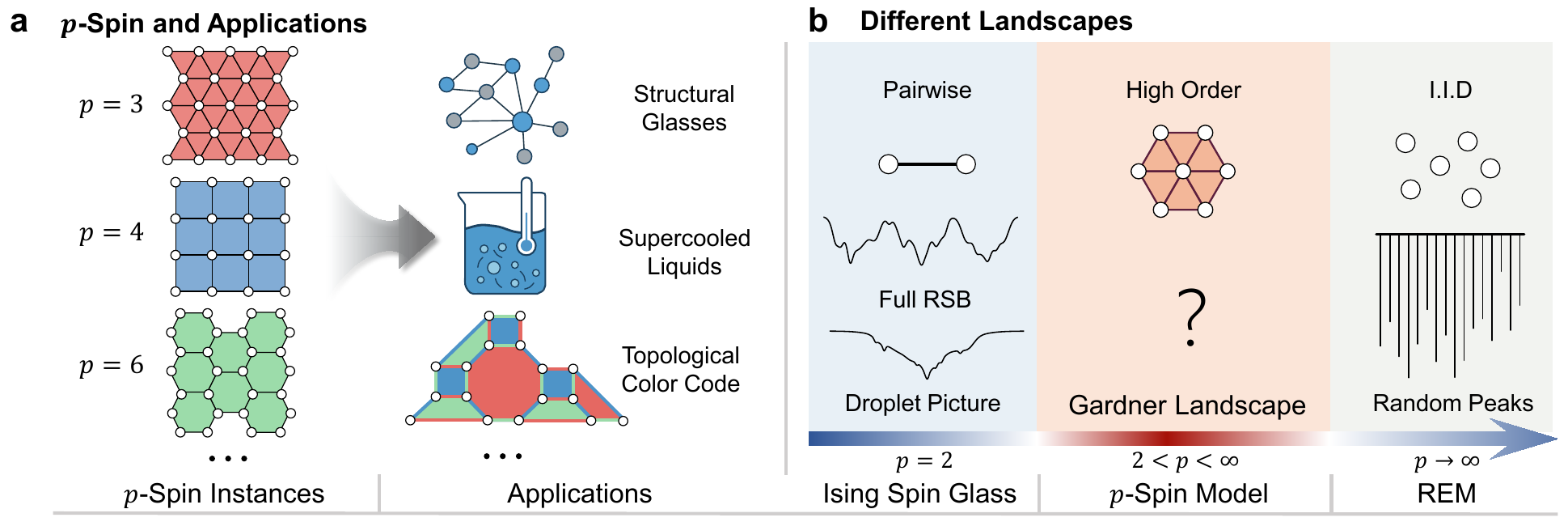}
\caption{Structural topologies and landscape complexity of $p$-spin glasses.
\textbf{a,} Depiction of higher-order structural topologies and their broad physical applications. Left: Representative $p$-spin models on triangular ($p=3$), square ($p=4$), and hexagonal ($p=6$) Edwards–Anderson lattices. Right: These models offer theoretical tools for probing complex systems such as structural glasses and supercooled liquids, and for tackling decoding problems in quantum topological color codes. \textbf{b,} Hierarchy of energy landscape topologies across $p$-spin models. For $p=2$ (SK model), the landscape reflects the tension between the full replica symmetry breaking (RSB) scenario and the droplet picture. In the intermediate range $2 < p < \infty$, the system enters a Gardner phase, creating a rugged, fractal‑like free‑energy landscape. In the limit $p \to \infty$, the landscape reduces to the uncorrelated energy levels of the Random Energy Model (REM).}
\label{fig:case_study_ab}
\end{figure*}

Unlike the well‑studied pairwise case ($p=2$), instances with $p\geq3$ give rise to a rugged, fractal‑like free‑energy landscape\cite{gardner_spin_1985,charbonneau_fractal_2014} (Fig.~\ref{fig:case_study_ab}) and are notoriously resistant to both exact and heuristic solvers. Exact algorithms (branch‑and‑bound\cite{hartwig1984recursive}, branch‑and‑cut\cite{de1995exact}) are limited to tens of spins, while metaheuristics such as simulated annealing (SA)\cite{grest1986cooling} and parallel tempering (PT)\cite{earl2005parallel} suffer from slow mixing and frequent entrapment in local minima and often require an impractically large number of sweeps to locate low‑energy configurations.

Recently, machine learning has emerged as a powerful paradigm for combinatorial optimization\cite{khalil2017learning,kwon2020pomo,li2018combinatorial,fan2020finding}, particularly through learning-based models. In physical and energy-based settings, notable examples include physics-inspired graph neural networks (PIGNN) \cite{schuetz2022combinatorial}, hypergraph-oriented optimization frameworks like hypOp \cite{heydaribeni2024distributed}, and the Free-Energy Machine (FEM) \cite{shen2025free}. These approaches achieve strong results on specific problem classes by embedding energy minimization or landscape navigation into gradient‑based learning. Yet these approaches are tightly coupled to particular task formulations and lack a unified training‑inference pipeline, limiting their cross‑domain applicability. An alternative line of work frames optimization as sequential decision‑making and leverages reinforcement learning (RL) to construct solutions step by step. DIRAC\cite{fan2023searching}, for instance, learns efficient spin‑flip policies for Edwards–Anderson spin glasses and achieves state‑of‑the‑art performance for pairwise ($p=2$) interactions. However, extending such RL‑based solvers to arbitrary higher‑order couplings has remained an open challenge. This limitation is not merely incremental: interactions of order $p\geq3$ create many‑body correlations that are exponentially harder to model and require fundamentally different state representations and credit assignment mechanisms.

In this work, we introduce \model~(\textbf{\underline{P}}-spin-g\textbf{\underline{LA}}ss model optimization leveraging deep rei\textbf{\underline{N}}for\textbf{\underline{C}}ement learning and hypergraph neural networ\textbf{\underline{K}}s), a reinforcement learning framework that, for the first time, solves the ground‑state problem for $p$-spin glasses with arbitrary interaction order $p$ in a unified manner. \model~is built on three key innovations. First, it directly operates on the native hypergraph representation of $p$-spin Hamiltonians, eliminating the need for auxiliary variables or quadratization and bypassing artificial pairwise reductions. The specially designed hypergraph neural network (PHGNN) encodes spin states and many‑body couplings into order-independent features, enabling \model~to seamlessly scale to higher orders. Second, \model~systematically exploits gauge symmetry\cite{nishimori1983gauge}—a fundamental property of spin glasses—throughout both training and inference. This symmetry‑aware design drastically reduces the search space, accelerates training convergence, and enhances final solution quality. Third, \model~provides a unified solver that, once trained on small synthetic instances, transfers zero‑shot to substantially larger systems and to a wide spectrum of NP‑hard combinatorial problems, including random $k$-XORSAT, hypergraph max‑cut, and conventional max‑cut—without any task‑specific customization.

\section*{PLANCK architecture and empirical performance}
\subsection*{Reinforcement learning formulation}
\begin{figure*}[t]
\centering
\includegraphics[width=0.94\textwidth]{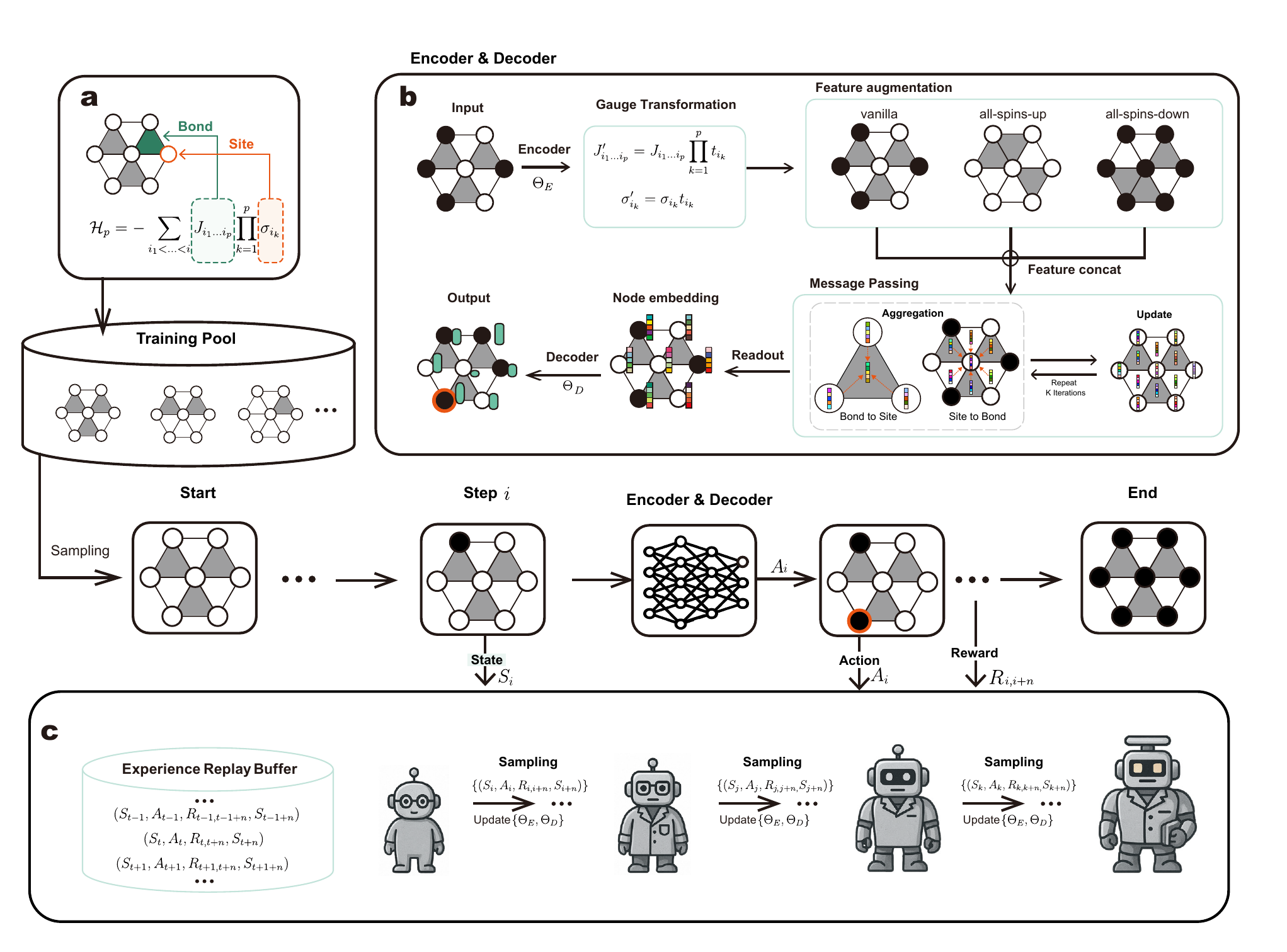}
\caption{\model~framework (I): training stage and learning components.
During training, we generate synthetic small $p$-spin glass instances and represent each instance as a hypergraph (nodes correspond to sites and hyperedges represent multi-spin interactions), which are added to the training pool. The \model~agent is optimized by sampling from this pool and interacting with instances in episodes that traverse from an all-spins-up to an all-spins-down configuration. At each step, gauge transformations produce energetically identical representations; the encoder $\Theta_\mathcal{E}$ performs message passing to produce node embeddings, and the decoder $\Theta_\mathcal{D}$ estimates $Q$-values for spin flips. Experience transitions are stored in a replay buffer and mini-batches are sampled to update $\{\Theta_\mathcal{E}, \Theta_\mathcal{D}\}$ via gradient descent.}
\label{fig:planck_framework_a}
\end{figure*}

We formulate the $p$-spin ground-state search as a Markov decision process (MDP), specified by the tuple $(S, A, R, P, \gamma)$, on the problem’s natural hypergraph $\mathcal{G} = (\mathcal{V}, \mathcal{E})$ (Fig.~\ref{fig:planck_framework_a}). In this representation, nodes $v_i \in \mathcal{V}$ (sites) correspond to spin variables $\sigma_i$, while hyperedges $e \in \mathcal{E}$ (bonds) encode the $p$-body couplings $J_e = J_{i_1, i_2, \dots, i_p}$. This setup allows the agent to learn a sequential spin-flip policy that directly minimizes the Hamiltonian, avoiding quadratic reformulations.

A state $s_t \in S$ encodes the spin configuration $\boldsymbol{\sigma}^{(t)} = \{ \sigma_1^{(t)}, \dots, \sigma_N^{(t)} \}$. An action $a_t \in A$ corresponds to the specific spin selected for flipping, inducing a transition to $s_{t+1}$ with probability $P(s_{t+1} \mid s_t, a_t)$. The immediate reward $r(s_t, a_t, s_{t+1}) \in R$ measures the resulting energy reduction and is given by
$
r(s_t, a_t, s_{t+1}) = 2 \sigma_{a_t} \sum_{e \in \mathcal{E}(a_t)} J_e \prod_{k \in e \setminus \{a_t\}} \sigma_k,
$
where $\mathcal{E}(a_t) = \{ e \in \mathcal{E} \mid a_t \in e \}$ is the set of interactions involving the flipped spin. As it is computed analytically from the current configuration and the coupling tensor, this reward is both computationally efficient and statistically unbiased. Future rewards are discounted by a factor $\gamma$.

To navigate the non-convex and rugged energy landscape characteristic of $p$-spin glass systems, we deliberately restrict each episode to start at the all-spins-up configuration and terminate at the all-spins-down configuration. This constraint drastically reduces the number of possible trajectories, which would compromise global exploration in a naive implementation. We resolve this tension by exploiting gauge symmetry (Eq. (2)), which maps any configuration to the all-up state while preserving the energy. This insight leads to a dual use of gauge transformation: feature-level augmentation during training and instance-level path resetting during inference. The resulting framework explores the configuration space efficiently while retaining the sample-efficiency benefits of a fixed start-end pair.

\subsection*{PLANCK architecture}
\begin{figure*}[t]
\centering
\includegraphics[width=0.94\textwidth]{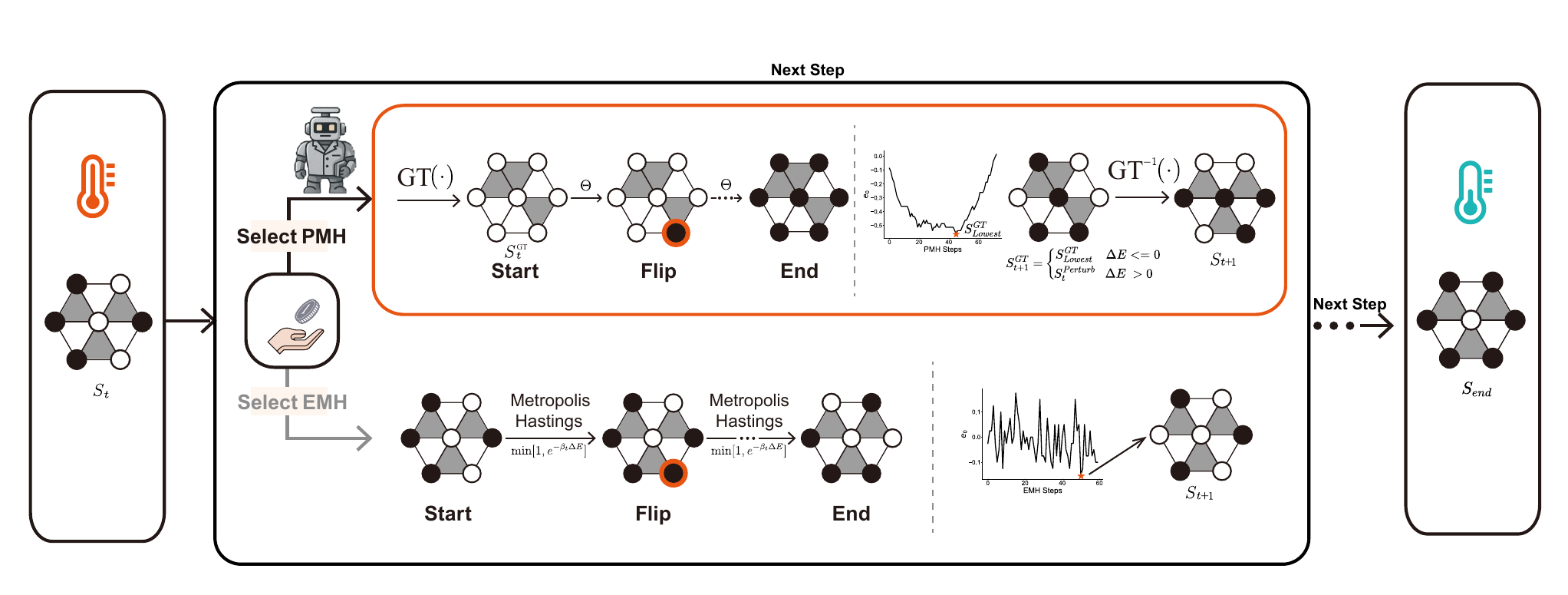}
\caption{\model~framework (II): application stage and hybrid inference.
In the application stage, an instance is initialized at a high-temperature phase with a random configuration $S_t$. The system probabilistically selects between a traditional energy-based metaheuristic method (EMH) and a \model~metaheuristic (PMH). If PMH is chosen, a gauge transformation GT($\cdot$) maps the system to an equivalent representation; the trained \model~performs coordinated spin flips to produce an energy trajectory, and a conditional update mechanism is applied (updating to the lowest-energy state when $\Delta E \le 0$, or to a perturbed state when $\Delta E > 0$), followed by the inverse transformation GT$^{-1}$($\cdot$) to yield the next state. If EMH is selected, a Metropolis-Hastings annealing process is executed, accepting flips with probability $\min[1, e^{-\beta_t \Delta E}]$. The global minimum-energy configuration found through this hybrid strategy is output as the predicted ground state in the low-temperature phase.}
\label{fig:planck_framework_b}
\end{figure*}

Given the intricate challenge of optimizing $p$-spin models, we develop \model~ to learn the policy network $\pi_\Theta$. During training (Fig.~\ref{fig:planck_framework_a}), \model~interacts online with randomly initialized small-scale $p$-spin glass instances, producing experience tuples $(s_t,a_t,r_t,s_{t+1})$ that are stored in a replay buffer $\mathcal{B}$ and periodically sampled to update the parameters $\Theta$ via gradient descent on the temporal-difference error.
During the testing stage (Fig.~\ref{fig:planck_framework_b}), \model~ combines its trained $Q$ network with simulated annealing for optimized $p$-spin glass solutions. The system alternates between neural-guided spin flips (based on $Q$ values) and thermal sampling, dynamically adjusting this balance via a temperature-dependent selection probability.

The success of \model~ largely depends on two key elements: (1) How can the $p$-spin glass system be represented, taking into account its physical characteristics? (2) How can these representations be used to compute a $Q$-value that predicts the long-term benefits of an action in a given state? We describe these challenges as the encoding and the decoding problem, respectively.

We design an encoder based on hypergraph neural networks \cite{feng2019hypergraph,yadati2019hypergcn,dong2020hnhn} named \ENCmodel~ ($p$-spin HyperGraph Neural Network), a gauge symmetry-aware Message Passing Neural Network \cite{gilmer2017neural} to capture the complex high-order interactions in the $p$-spin glass system. \ENCmodel~begins with carefully designed features. For each site, we design its feature as $\mathbf{x}_s = (\sigma_i, h_i) \in \mathbb{R}^2$, where $\sigma_i \in \{-1, +1\}$ indicates the current state of the site and $h_i = \sum_{e \in \mathcal{E}(i)} J_e \prod_{k \in e \setminus \{i\}} \sigma_k$ represents the local field contribution from $p$-body interactions, with $\mathcal{E}(i)$ denoting the set of hyperedges incident to site $i$. For each bond, we design its feature as $x_b = J_{i_1 \dots i_p} \in \mathbb{R}^1$, where $J_{i_1 \dots i_p}$ is the coupling strength from the Hamiltonian. These features naturally decouple feature complexity from the interaction order $p$, allowing the model to handle higher-order interactions without explosion of the feature space. We note that the Hamiltonian of the $p$-spin glass system is invariant between the instances $(J_{i_1,i_2, \dots i_p},\sigma_{i_k})$ and $(J_{i_1,i_2, \dots i_p}',\sigma_{i_k}')$ through the technique of gauge transformation (GT) \cite{wegner1971duality,nishimori1981internal,ozeki1995gauge}:
\begin{equation}\label{eq:p-spin-gt}J_{i_1,i_2, \dots i_p}' = J_{i_1,i_2, \dots i_p}\prod_{k=1}^p t_{i_k}, \quad \sigma_{i_k}' = \sigma_{i_k}t_{i_k}
\end{equation}
where the gauge variables $t_{i_k} \in \{-1,+1\}$ induce symmetry transformation which is capable of toggling the spin glass system between any two configurations while keeping the system energy invariant. We use a simple but effective feature augmentation trick to convert the raw features $(\mathbf{x}_s, \mathbf{x}_b)$ into gauge-equivalent augmented features $(\mathbf{x}_s^{\text{(aug)}}, x_b^{\text{(aug)}})$:
\begin{equation}\label{eq:gt-concat}
\begin{cases}
\mathbf{x}_s^{\text{(aug)}} = [\mathbf{x}_s, \mathbf{x}_s^{\text{(up)}}, \mathbf{x}_s^{\text{(down)}}] \in \mathbb{R}^{6} \\[5pt]\mathbf{x}_b^{\text{(aug)}} = [x_b, x_b^{\text{(up)}}, x_b^{\text{(down)}}] \in \mathbb{R}^3
\end{cases}
\end{equation}
where $(\mathbf{x}_s^{\text{(up)}}, x_b^{\text{(up)}})$ and $(\mathbf{x}_s^{\text{(down)}}, x_b^{\text{(down)}})$ represent the features derived through a gauge transformation that sets the system to an all-spins-up configuration and an all-spins-down configuration, respectively. This feature augmentation integrates gauge invariance directly within the embedding process. The features $(\mathbf{x}_s^{\text{(aug)}}, x_b^{\text{(aug)}})$ are then processed through a two-stage message passing process which alternates between sharing information from site-to-bond and bond-to-site aggregations (see Figs.~\ref{fig:planck_framework_a} and \ref{fig:planck_framework_b}). Once message passing is completed, we get the site embeddings $\mathbf{z}_i$ that encapsulate both individual conditions and their shared surroundings within $K$ hops. These are aggregated by summation to produce a global system representation $\mathbf{z}_{\text{s}}$, which is invariant to permutation and maintains gauge symmetry. Next, we use $\mathbf{z}_i$ to represent states and $\mathbf{z}_{\text{s}}$ to denote the global context in the Markov Decision Process of \model.

For decoding, we design a deep parameterization of the score function (the $Q$-function) 
which operates on learned representations of states and actions, denoted as $\mathbf{z}_s$ and $\mathbf{z}_a$ respectively, to compute the state-action value function $Q(s, a^{(i)}; \Theta)$. This function estimates the expected future cumulative rewards associated with action $a^{(i)}$ in state $s$, while following the policy $\pi_\Theta(a^{(i)} \mid s)$ throughout the entire episode. The $Q$-function takes the following form:
\begin{equation}\label{equ:qnet}
Q(s, a^{(i)}; \Theta) = \text{MLP}([\mathbf{z}_s,\mathbf{z}_i];\Theta)
\end{equation}
Here MLP denotes a two-layer multi-layer perceptron network with ReLU activation functions, and $\Theta = \{\Theta_E, \Theta_D\}$, where $\Theta_E$ denotes the encoder parameters and $\Theta_D$ represents the decoder parameters.

The training pipeline of \model~ uses $n$-step $Q$-learning to fine-tune the $Q$-function parameters $\Theta$ for optimizing the $p$-spin model. Experience transitions are stored in a replay buffer $\mathcal{B}$, with each transition denoted by a tuple $(s_t, a_t, r_{t,t+n}, s_{t+n})$. The main objective of the training is to minimize the $n$-step $Q$-learning loss, which measures the difference between predicted $Q$-values and $n$-step bootstrapped targets. This loss function combines immediate rewards with discounted future rewards via n-step reward accumulation, with the discount factor $\gamma$ adjusting the significance of future rewards. The formula for the $n$-step $Q$-learning loss function is given as:
\begin{linenomath*}
\begin{equation}
    \begin{aligned}
    \mathcal{L} = {} & \mathbb{E}_{(s_t, a_t, r_{t,t+n}, s_{t+n}) \sim \mathcal{B}} \Bigl[
    \bigl( r_{t,t+n} + \gamma \max_{a_{t+n}} Q(s_{t+n}, a_{t+n}; \hat{\Theta}) \\
    & \qquad\qquad\qquad\qquad - Q(s_t, a_t; \Theta) \bigr)^2 \Bigr].
    \end{aligned}
\end{equation}
\end{linenomath*}

We periodically draw mini-batches from the replay buffer $\mathcal{B}$, where each sample consists of a 4-tuple $(s_t, a_t, r_{t,t+n}, s_{t+n})$, and use gradient descent to refine the parameters $\Theta$. The target network parameters $\hat{\Theta}$ are maintained separately and periodically synchronized with $\Theta$ at fixed intervals to ensure training stability.

\model~is trained on a large number of small-scale random $p$-spin glass instances. Once trained, it serves as a solver that can be applied to efficiently and repeatedly search for ground states of $p$-spin glass instances that are significantly larger than those seen during training. Similar to the training process, it begins from the all-spin-up configuration and proceeds to the all-spin-down configuration, flipping at each step the spin with the highest $Q$-value as predicted by the trained model. Notably, \model~always starts from the same initial configuration, which helps drastically reduce the number of potential trajectories and reduces the amount of training data required. Ending at a fixed terminal configuration ensures that the agent completes the MDP within a finite number of steps. This constraint enables the agent to learn an efficient, data-driven spin-flipping strategy with minimal regret.

However, we emphasize that while the strategy adopted by \model~provides significant efficiency gains, it also introduces certain limitations. Specifically, starting from a fixed initial state and proceeding to a fixed terminal state, each intermediate configuration is explored exactly once. Although this single-pass exploration is highly likely to succeed, it severely restricts the agent's exploration capability, particularly in extremely large decision spaces. In contrast, heuristic methods such as simulated annealing owe much of their success to the ability to initiate from diverse states and revisit promising configurations multiple times through stochastic perturbations. Therefore, to further enhance the agent’s ability to locate ground states, one straightforward direction is to endow \model~with similar capabilities. Unfortunately, the current strategy adopted by \model~does not permit such flexible exploration.

To address the aforementioned limitation, we adopt a simple yet effective technique known as gauge transformation (GT), applied at the instance level, similar to its use at the feature level. After the agent reaches the final spin configuration in each trajectory, we apply a GT that resets the spin configuration back to the initial all-spin-up configuration. This transformation simultaneously modifies the bond weights such that the total energy of the system remains invariant. In doing so, GT enables the agent to restart exploration from a consistent initial configuration, thereby extending the effective search horizon without violating energy conservation principles.

Through this mechanism, we introduce a hybrid approach for solving challenging spin systems by integrating the predictive capabilities of \model~with the stochastic exploration of simulated annealing (SA). At each temperature $\beta_t$ in the annealing schedule, the system probabilistically selects between traditional Metropolis-Hastings updates and \model-guided spin flips:
\begin{equation}
P_{\text{select}} = \begin{cases}
\text{SA: } \min[1, e^{-\beta_t \Delta E}] & \text{with probability } p \\
\text{\model: } Q\text{-value optimization} & \text{with probability } 1 - p
\end{cases}
\end{equation}

If the spin flip suggested by \model~fails to lower the energy (i.e., $\Delta E \geq 0$), thermal perturbations are introduced with probability:
\begin{equation}
P_{\text{perturb}} = q \times \frac{\beta_{\max} - \beta_t}{\beta_{\max} - \beta_{\min}},
\end{equation}
where $q = 0.5$ balances the trade-off between exploration and exploitation. This adaptive blending allows SA to dominate at high temperatures ($\beta_t \rightarrow \beta_{\min}$) for global exploration, while favoring \model's learned $Q$-value gradients at low temperatures ($\beta_t \rightarrow \beta_{\max}$) for precise local optimization.

During testing, this hybrid framework demonstrates superior performance compared to conventional SA. While traditional SA relies purely on local energy differences $\Delta E$, \model's Q-network encodes global multi-spin correlations through learned parameters $\Theta$. The GT-based reset mechanism allows periodic non-local transitions via:
\begin{equation}
(\boldsymbol{\sigma}_{\text{up}}^{(s)}, J^{(gt)}) = \text{GT}(\boldsymbol{\sigma}^{(s-1)}, J)
\end{equation}
preserving the system's topology while enabling ergodic traversal of gauge-equivalent configurations. The temperature-adaptive perturbation probability $P_{\text{perturb}}$ provides a smooth interpolation between thermal and learned behaviors, enhancing the ability to tunnel out of local minima. 

\subsection*{Training details and baseline solvers}
We selected \model~hyperparameters via ablation studies, focusing on hypergraph neural networks and reinforcement learning aspects such as message-passing layers $k$, RL discount $\gamma$, and reward horizon $n$. While \model's current performance is already robust, we note that additional improvements could potentially be achieved through more systematic hyper-parameter optimization. In particular, we found that the optimal number of message-passing layers $k$ depends significantly on both the system size and interaction order $p$, with $K=k$ performing best for most cases we tested. This finding partially confirms our initial hypothesis that deeper message passing would better capture complex correlations.

For baselines, we implemented two classical optimization techniques. For SA, per Ref.~\cite{wang_comparing_2015}, we utilized a linear temperature annealing from $\beta_{\min}=0$ to $\beta_{\max}=5$ with $N_t=100$ steps and $N_s=50$ sweeps each. For PT, as guided by Ref.~\cite{roma_ground_2009}, we used $N_r=20$ replicas across $T\in[0.1,1.6]$, swapping configurations every $N$ flips.

The greedy search algorithm optimizes energy locally by flipping individual spins. It stops when further flips cannot reduce energy, settling in a local minimum. Though efficient, it often stalls in suboptimal states due to the spin glasses' intricate energy landscape.

\subsection*{Optimizing the $p$-spin models}
To systematically evaluate \model's capability in optimizing $p$-spin glass systems with different sizes $L$, we conducted extensive numerical experiments on both Gaussian and bimodal bond weight distributions across multiple lattice geometries.  We first trained \model~on small-scale synthetic systems: for triangular ($p=3$) and square ($p=4$) lattices we set $L=5$, while for the hexagonal lattices ($p=6$) we set $L=4$. All training and testing datasets are generated by the lattice graph generation functions from the Networkx package \cite{hagberg2020networkx}. Subsequently, we tested the trained models on larger system sizes to evaluate their generalization capability. For baseline comparisons, we selected two classical approaches for spin glass problems: Simulated Annealing (SA) \cite{kirkpatrick1983optimization} and Parallel Tempering (PT) \cite{earl2005parallel}. Since all methods (\model, SA, and PT) share a similar inner loop structure where spins are flipped from a special initial spin configuration, the total computational budget for each algorithm is naturally quantified by the number of distinct initial configurations processed, which we denote as $N_{\text{initial}}$.

\begin{figure*}[t]
\centering
\includegraphics[width=1.0\textwidth]{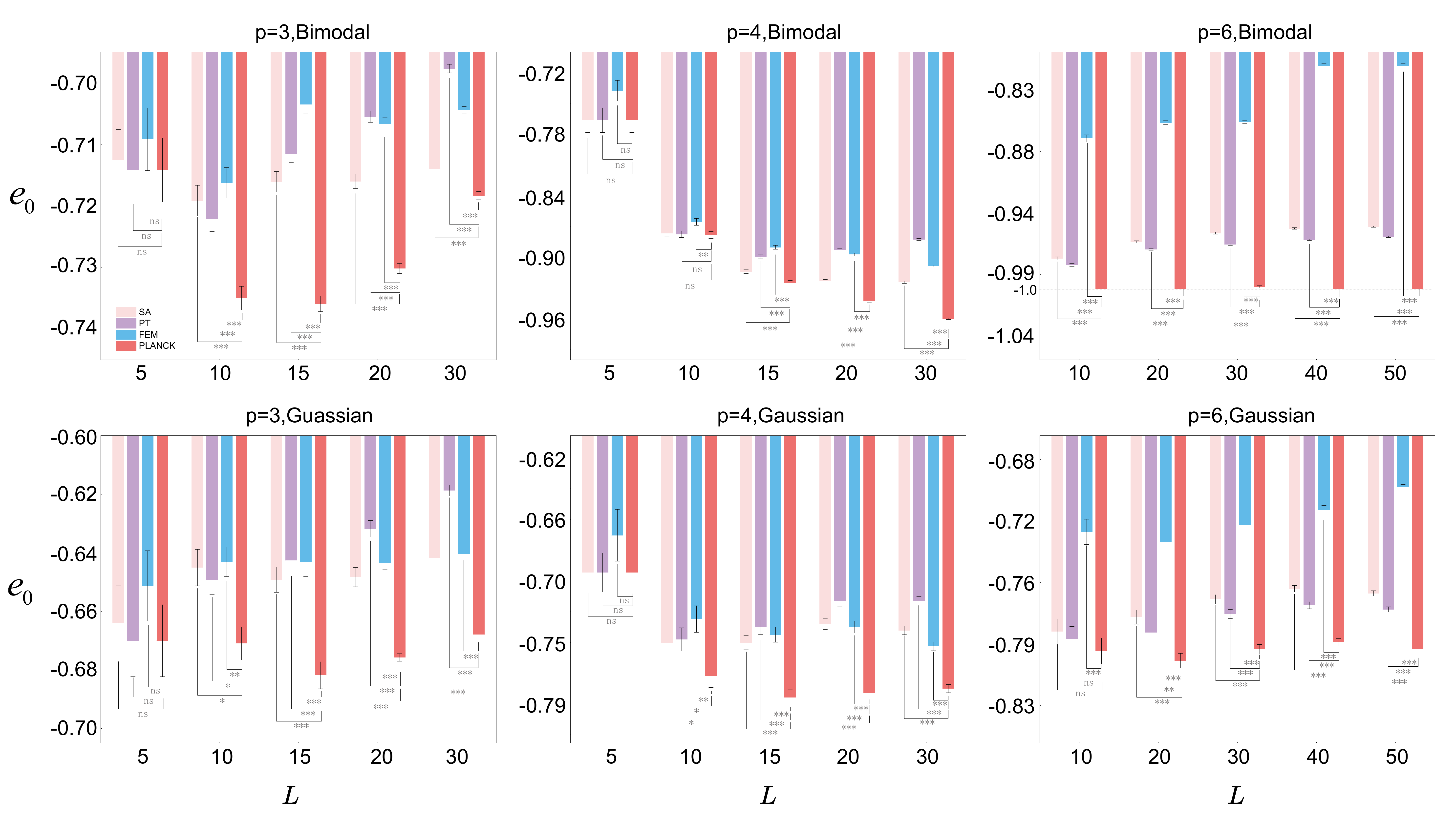}
\caption{Performance of different methods in minimizing the $p$-spin Hamiltonian.
We compared the disorder averaged ``ground-state'' energy per bond (predicted by each method), denoted as $e_0$, to benchmark various methods. Each result is computed over 50 independent instances, with mean values (bars) and standard error of the mean (SEM; error bars) reported. Note that \model~tested here is trained exclusively on small synthetic instances ($L=5$ for triangular $p=3$ and square $p=4$; $L=4$ for hexagonal $p=6$). It is then tested, without any retraining or fine‑tuning, on systems 4–6 times larger than seen during training. Both simulated annealing (SA) and parallel tempering (PT) use identical computational budget ($N_{\rm init}=5000$). It can be observed that across all lattice types and both Bimodal (a–c) and Gaussian (d–f) couplings, \model~(red) consistently finds the lowest‑energy configurations among all other methods. The presented results demonstrate that the policy \model~learned on small instances transfers effectively to much larger, never‑before‑seen instances.}
\label{fig:more_results_in_finding_ground_state}
\end{figure*}

To ensure fair comparison, we fixed $N_{\text{initial}}=5000$ for all methods. For SA, this budget is allocated across $N_T$ temperature steps with $N_S$ initial configurations processed at each step ($N_{\text{initial}} = N_s \times N_t$). The PT implementation distributes the budget over $N_e$ epochs of $N_r$ replicas, with each epoch-replica combination processing one initial configuration ($N_{\text{initial}} = N_e \times N_r$). Fig.~\ref{fig:more_results_in_finding_ground_state}(a-c) present the comparative performance for bimodal coupling distributions across the three lattice types on 50 independent instances respectively, with results reported as mean$\pm$SEM. The results demonstrate \model's superior solution quality, with particularly notable advantages emerging at larger system sizes. This scaling behavior provides strong evidence for \model's enhanced generalization capability compared to the baseline methods. The performance advantage remains consistent for Gaussian coupling distributions as shown in Fig.~\ref{fig:more_results_in_finding_ground_state}(d-f), with \model~maintaining significantly lower final energies across all tested system sizes and lattice geometries. 

For small $p$-spin models, exact solutions can be obtained using branch-and-bound solvers such as Gurobi \cite{gurobi}. However, as the system size $L$ increases, finding the ground state with exact methods becomes computationally intractable. Heuristic or approximate algorithms like SA, PT, and \model~are impacted by the parameter $N_{\text{initial}}$, which is the count of independent search attempts. Intuitively, larger values of $N_{\text{initial}}$ generally lead to better performance, although at the cost of increased computation time. To investigate how $N_{\text{initial}}$ affects algorithm performance, we conducted experiments on three EA-$p$ spin models with Gaussian couplings: $p=3$ ($L=20,25,30$), $p=4$ ($L=15,20,30$) and $p=6$ ($L=10,16,20$). We produced 50 independent test cases for each scenario, altering $N_{\text{initial}}$ between 1 and $2\times10^4$ to investigate the variation in $e_0$ (energy density per bond) as $N_{\text{initial}}$ rises. For PT, we recorded $e_0$ at each increment of $N_{\text{initial}}$, while for SA and \model, each data point represents results using 5000 initial configurations.

Fig.~\ref{fig:more_results_in_finding_ground_state} shows the mean energy density $e_0$ with SEM (shaded areas) for all algorithms across the test sets. Several key observations emerge from these results. First, \model~consistently achieves the best performance at any given $N_{\text{initial}}$ value, demonstrating statistically significant advantages over the baseline methods. Notably, \model's performance with just 5000 initial configurations surpasses that of both SA and PT even when they use $2\times10^4$ initial configurations. Finally, the trends with increasing $N_{\text{initial}}$ reveal important differences between the algorithms. PT (blue curves) shows the most pronounced performance improvement as $N_{\text{initial}}$ increases, with its energy density decreasing substantially across the tested range. SA exhibits more moderate improvements with additional initial configurations (brown curves). In contrast, \model's performance remains remarkably stable regardless of $N_{\text{initial}}$ (red curves), indicating that its carefully designed reinforcement learning components and hypergraph representation learning modules enable highly efficient search without requiring extensive sampling of initial states. 

Beyond the effectiveness, \model~ achieves exceptional computational efficiency through carefully designed algorithmic strategies. The framework employs a dynamic hybrid approach during inference that strategically balances neural-guided spin flips with traditional Monte Carlo sweeps, which enables precise control over the speed-accuracy trade-off.

All experiments were conducted on a 64-core computer with 512GB RAM and a NVIDIA A100 GPU with 32GB memory. Although \model~ requires an initial offline training phase, this cost is amortized as the trained model can be applied indefinitely to systems of varying sizes within the same lattice type without retraining. The computational complexity of \model~ during the testing stage scales favorably with system size. For sparse hypergraph systems, the time complexity remains linear $\mathcal{O}(N)$ due to efficient message passing in the hypergraph neural network, while dense systems exhibit quadratic scaling $\mathcal{O}(N^2)$. The gauge-invariant feature engineering requires only $\mathcal{O}(N + E)$ operations, processing each spin and hyperedge in parallel. The encoding phase, implemented through $K=5$ sparse message-passing layers, maintains $\mathcal{O}(KNd^2)$ complexity for typical systems by leveraging hypergraph sparsity. The lightweight decoder contributes minimal overhead with $\mathcal{O}(Nh^2)$ complexity, where $h=32$.

Notably, the incorporation of gauge transformations provides significant training acceleration, achieving faster convergence compared to non-invariant baselines while simultaneously improving the final approximation ratio. This improvement stems from the constrained policy search space enabled by symmetry-aware representations. These advances establish \model~ as an efficient and scalable framework for solving complex high-order optimization problems in both physics and computer science domains.

\subsection*{Applications on General NP-hard problems}
\begin{figure*}[t]
\centering
\includegraphics[width=1.0\textwidth]{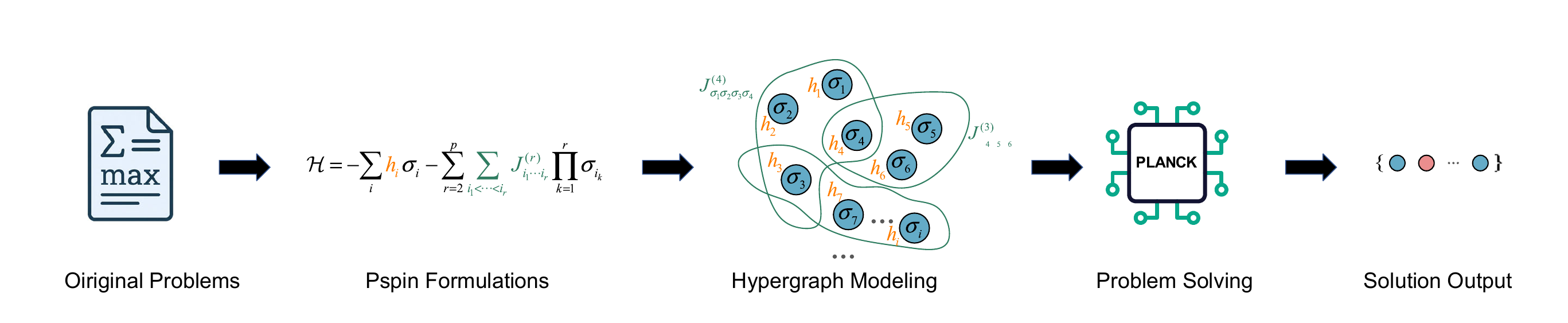}
\caption{\model~ as a unified solving framework for NP-hard problems (workflow).
The framework unifies the optimization workflow by recasting general NP-hard combinatorial optimization problems, originally expressed via objectives $\mathcal{F}$, into a generalized $p$-spin Hamiltonian $\mathcal{H}$. This Hamiltonian is encoded as a hypergraph, on which the \model~ iteratively optimizes the spin configuration to find the ground state, subsequently translating it back into a solution of the initial problem.}
\label{fig:planck_applications_a}
\end{figure*}

\begin{figure*}[t]
\centering
\includegraphics[width=1.0\textwidth]{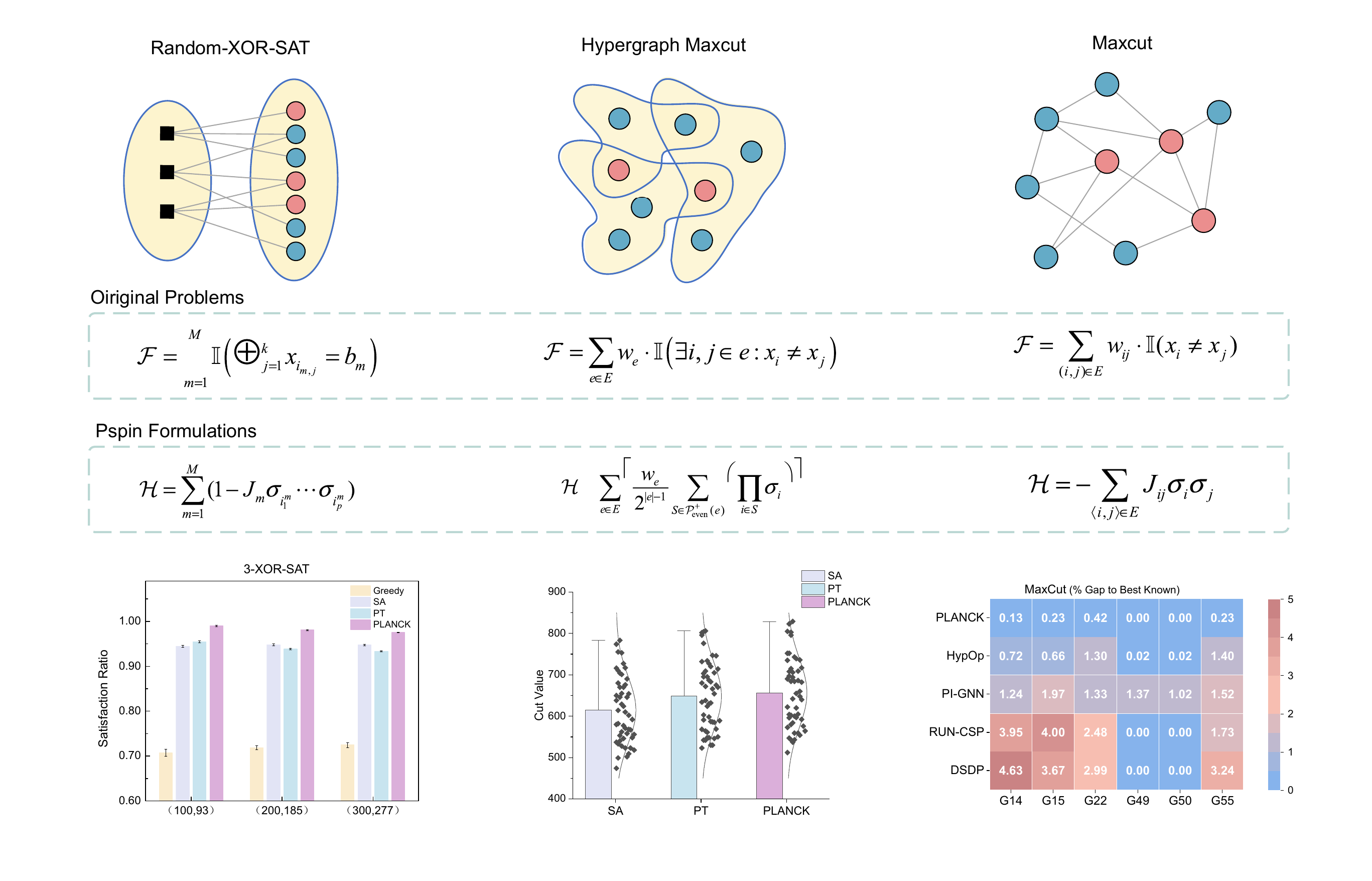}
\caption{\model~ as a unified solving framework for NP-hard problems (case studies).
Empirical evaluations across three distinct representative problems demonstrate the \model's versatility and performance. For each problem, the original objective function and its corresponding $p$-spin formulation are explicitly shown. In the Random 3-XOR-SAT benchmarks, comprising 50 synthetic instances per problem scale generated via the Python \texttt{cnfgen} library, \model~maintains a near-optimal satisfaction ratio across varying problem scales (e.g., $N=100, 200, 300$), significantly outperforming Greedy, Simulated Annealing (SA), and Parallel Tempering (PT) baselines while remaining robust to system size growth. For the Hypergraph MaxCut problem, evaluations were conducted on a dataset of 50 complex instances, each comprising 50 nodes and 600 to 900 mixed-order couplings (including interactions such as $p=2$ and $p=4$). \model~achieves higher mean cut values and lower variance compared to SA and PT, demonstrating its capacity to effectively navigate highly rugged, multi-body energy landscapes. On the classical MaxCut problem using standard Gset benchmarks, \model~attains the smallest relative gap to the best-known results compared to other advanced heuristic and learning-based methods (Hypop, PI-GNN, RUN-CSP, DSDP). It exactly matches the optimal cuts for several instances (e.g., G49, G50), reflecting strong generalization capabilities to diverse, real-world graph structures.}
\label{fig:planck_applications_b}
\end{figure*}

The Ising formulation has emerged as a universal paradigm for representing combinatorial optimization problems, capable of encoding Karp's 21 NP-complete problems \cite{lucas2014ising}. Although higher-order Ising models can be reduced to quadratic form \cite{rosenberg1975reduction}, this typically requires auxiliary variables, which can inflate problem size and computational cost \cite{mandal2020compressed,glos2022space,bybee2023efficient}. Consequently, there is growing interest in tackling higher-order ($p$-spin) interactions directly in their native form \cite{bybee2023efficient,bhattacharya2024computing,nikhar2024all,pedretti2025solving}, thereby avoiding the overhead of variable expansion. To evaluate \model's ability to solve high-order NP-hard problems, we design an end-to-end framework that first transforms original problems into $p$-spin models. The \model~is then trained to find their ground states, which are mapped back to solutions for the original optimization tasks. An overview of the end-to-end workflow is shown in Fig.~\ref{fig:planck_applications_a}, and representative case studies are summarized in Fig.~\ref{fig:planck_applications_b}.

Specifically, we categorize $p$-spin models transformed from NP-hard problems into three types: pure $p$ problems with interactions of precisely $p$-spins (e.g., random $k$-XORSAT); mixture $p$ problems with varied interaction orders (e.g., hypergraph MaxCut); and pairwise Ising problems with solely quadratic terms (e.g., standard MaxCut).

Random $k$-XORSAT (pure $p$) is a key variant of Boolean satisfiability with wide applications in cryptography~\cite{massacci2000logical}, error-correcting codes~\cite{nandi2024margin,sourlas1989spin}, and hardware verification~\cite{gupta2006sat}. This problem aims to assign values to $N$ Boolean variables such that $M = cN$ parity constraints are satisfied. Each constraint is a $p$-tuple of random variable indices, associated with a binary value $y_m$. This problem can be translated to a $p$-spin model as follows:
\begin{equation}
    \mathcal{H}_{\text{random}-k-\text{XORSAT}} = \sum_{m=1}^M (1 - J_m \sigma_{i^m_1} \cdots \sigma_{i^m_p})
\end{equation}
Here $(1 - J_m \sigma_{i^m_1} \cdots \sigma_{i^m_p})$ denotes a parity condition, where $\sigma_{i^m_k} = (-1)^{x_{i^m_k}}$ reflects the original Boolean variable $x_{i^m_k}$ for the $k$-th variable in the $m$-th constraint. The $J_m = (-1)^{y_m}$ captures the desired parity $y_m$ for the $m$-th condition.

We conducted extensive experiments on both tough random 3-XORSAT and random 4-XORSAT problems. The results demonstrate that \model~achieves remarkable performance in solving these challenging instances.

Hypergraph Max-Cut (mixed $p$)~\cite{veldt2022hypergraph} extends the conventional Graph MaxCut problem and is of essential importance in combinatorial optimization with profound theoretical and practical impacts. Specifically, within a k-uniform hypergraph~\cite{kogan2015sketching}, where each hyperedge comprises precisely $k$ vertices, the problem is to partition the vertex set into two separate, non-empty groups to maximize the number (or total weight) of hyperedges linking these groups. This problem can subsequently be formulated as a $p$-spin model as follows~\cite{brakensiek2021mysteries}:
\begin{equation}
\mathcal{H}_{\text{Hypergraph MaxCut}} = \sum_{e \in E} \left[ \frac{w_e}{2^{|e|-1}} \sum_{S \in \mathcal{P}_{\text{even}}^+(e)} \left( \prod_{i \in S} \sigma_i \right) \right]
\end{equation}
Here $e$ denotes a hyperedge comprising $k$ vertices, and $w_e$ is the hyperedge's weight. The spin variables, $\sigma_i$ (where $\sigma_i \in \{-1, +1\}$), define the partitioning of the vertices. The set $\mathcal{P}_{\text{even}}^+(e) = \{ S \subseteq e \mid |S| \text{ is even}, S \neq \emptyset \}$ denotes the collection of all non-empty even-sized subsets of the hyperedge $e$. 

We evaluated \model's performance on $k$-uniform hypergraphs ($k=4,5$) of different sizes. The results demonstrate \model's proficiency in addressing hypergraph MaxCut problems. For 4-uniform hypergraphs, \model~surpassed SA and PT, especially with larger graphs. This advantage is even greater for 5-uniform hypergraphs.

Conventional MaxCut (pairwise $p$=2) is an NP-hard problem involving pairwise spin interactions and is crucial in fields such as operations research~\cite{alidaee19940}, computer vision~\cite{arora2010efficient}, and VLSI circuit design~\cite{barahona1988application}. For an undirected graph $G=(V,E)$ with $n=|V|$ vertices and edge weights $w_{ij}$, the MaxCut problem is to divide the vertex set $V$ into two non-overlapping subsets in a way that the total weight of the edges between them is maximized. Following the standard Ising model conversion~\cite{barahona1982computational}, we can transform this problem into a $p=2$ spin glass formulation:
\begin{equation}
\mathcal{H}_{\text{MaxCut}} = -\sum_{\langle i,j \rangle \in E} J_{ij}\sigma_i\sigma_j
\end{equation}
Here $\sigma_i \in \{-1,+1\}$ are Ising spins representing vertex partitions, $J_{ij} = \frac{w_{ij}}{2}$ are the couplings.

We train our model on synthetically produced Barabási-Albert (BA) graphs~\cite{barabasi1999emergence}, which replicate the scale-free characteristics typical of real-world networks. Then we evaluate \model's performance using the real-world Gset benchmark dataset~\cite{ye2003gset}. Our baseline includes heuristics methods such as DSDP~\cite{choi2000solving} and KHLWG~\cite{kochenberger2013solving}, as well as machine learning methods such as RUN-CSP~\cite{tonshoff2022one}, PI-GNN ~\cite{schuetz2022combinatorial}, and HypOp \cite{heydaribeni2024distributed}. The results demonstrate \model's remarkable effectiveness across all Gset benchmarks.

All the experiments above demonstrate that \model~can serve as a unified solver for combinatorial optimization problems, exhibiting strong performance across diverse settings. \model~not only generalizes across different scales (Fig.~\ref{fig:more_results_in_finding_ground_state}) but also efficiently adapts to various parameters and transfers between problem types. This robust generalization ability stems from \model's preservation of the underlying mathematical structure in $p$-spin glass representations, making it particularly suitable for real-world optimization tasks where problem specifications may vary dynamically.

\subsection*{Emergent human‑like strategy in the Baxter‑Wu model}
To systematically examine the outstanding performance of \model , we performed an in-depth interpretability analysis using the exactly solvable Baxter-Wu (BW) model \cite{baxter1973exact} as our test system. This canonical $p$-spin model ($p=3$), first solved exactly by Baxter and Wu fifty years ago \cite{baxter1973exact,baxter2016exactly}, has become a fundamental benchmark in the study of frustrated systems due to its rich theoretical properties and experimental relevance. The BW model ($p=3$) belongs to the same universality class as the two-dimensional four-state Potts model and the directional three-spin Ising model \cite{arashiro2003short}, despite their apparent structural differences. This equivalence makes it particularly valuable for understanding universal behavior in frustrated systems.

The Hamiltonian for the BW model on a triangular lattice is given by:
\begin{equation}
\mathcal{H} = -J \sum_{\langle i,j,k\rangle} \sigma_i \sigma_j \sigma_k
\end{equation}
Here $\sigma_i \in \{-1,+1\}$ represents Ising spins, $\langle i,j,k\rangle$ denotes the elementary triangular plaquettes, and $J$ is the uniform coupling constant that we set to 1 for our studies. The ground state of this system exhibits a fourfold degeneracy that reveals its underlying symmetry structure, consisting of one completely ferromagnetic state with all spins aligned along with three ferrimagnetic states where spins in two of the three sublattices are anti-aligned with the third. This semi-global symmetry, in which the Hamiltonian remains invariant under reversal of all spins in any two sublattices, plays a crucial role in the system's dynamics and provides an ideal testbed for understanding algorithmic behavior.

For our interpretability analysis, we initialize a 94-spin BW system with all spins in an anti-ferromagnetic configuration and without periodic boundary conditions to maximize frustration. We compare \model~against three baseline approaches: a greedy search algorithm that always flips the spin giving the maximum immediate energy reduction, SA with optimized parameters ($N_s=50$, $N_t=100$), and PT with optimized parameters ($N_e=16$, $N_r=20$). To isolate the effects of the reinforcement learning components, we configure \model~to use only its RL decision-making module without the SA augmentation.

\begin{figure*}[t]
\centering
\includegraphics[width=0.73\textwidth]{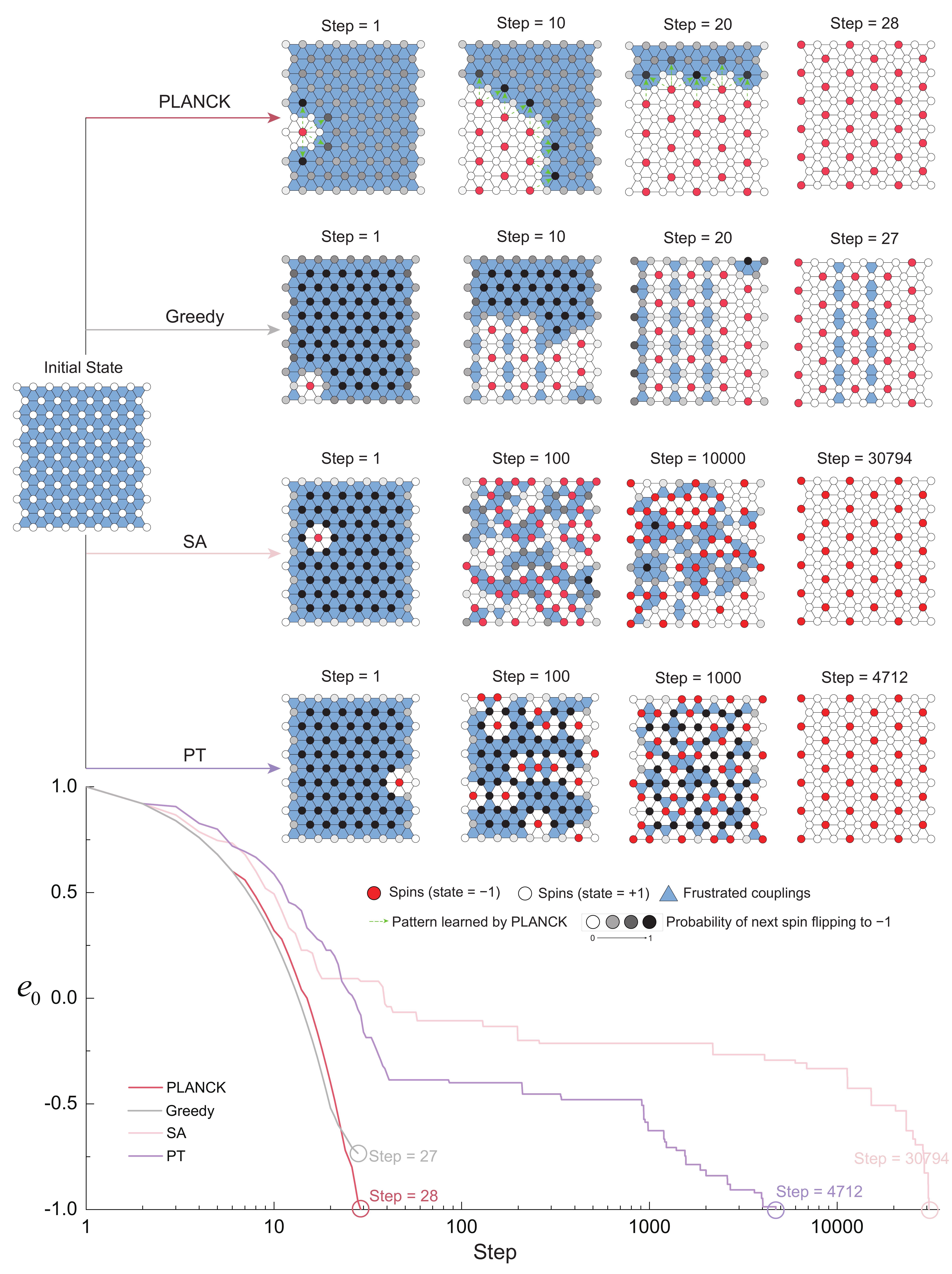}
\caption{\model~exhibits human-like intelligence by efficiently optimizing the ferromagnetic Baxter-Wu model with couplings $J=1$.
We evaluate various optimization methods in searching for the ground state of the Baxter-Wu model ($p=3$) with uniform ferromagnetic couplings ($J=1$) on a 94-spin triangular lattice without periodic boundary conditions. We compare \model~(using only its RL components) against Greedy search, SA, and PT, with all methods starting from the same maximally frustrated ferromagnetic configuration. The figure illustrates energy density changes (y-axis, 1 to -1) against optimization steps (x-axis). Crucial configurations are depicted at representative trajectory points, where empty circles denote $\sigma_i=-1$, filled circles signify $\sigma_i=+1$, and shaded triangles mark frustrated plaquettes. The PT results correspond to the lowest-energy replica. Besides, we highlight the configurations at key steps of the \model~algorithm, (\textbf{a})–(\textbf{g}) to illustrate the evolution of its frustration-resolving strategy during optimization. The results reveal different optimization behaviors: while Greedy search (27 steps) and \model~(28 steps) show comparable step efficiency, only \model~successfully reaches the ground state by employing an emergent strategy of flipping core spins in hexagonal clusters to simultaneously resolve multiple frustrations. In contrast, SA (30,794 steps) and PT (4,712 steps, shown for the lowest-energy replica) eventually find the ground state but through lengthy trajectories exhibiting random-walk characteristics without discernible pattern. The energy density trajectories emphasize \model's superior performance, achieving faster convergence and lower final energies than traditional methods, while keeping the optimization dynamics physically meaningful.}
\label{fig:explanation}
\end{figure*}

Fig.~\ref{fig:explanation} illustrates the entire evolution of the energy density and spin configurations.

Spins with $\sigma_i=-1$ and $\sigma_i=+1$ are depicted by empty and filled circles, respectively, and shaded triangles denote frustrated plaquettes. This comparative analysis reveals several key insights. Firstly, the greedy search algorithm, while aligned with \model~in step efficiency by design, is inherently unable to overcome local optima due to its localized optimization strategy. Secondly, analyzing the stochastic algorithms SA and PT, we notice that they eventually reach the ground state, albeit after exceedingly lengthy trajectories, specifically 30,794 and 23,836 steps. These pathways extend several orders of magnitude beyond the system size and lack identifiable patterns or physical interpretability. Although frustrated plaquettes reduce over time, this reduction occurs via seemingly arbitrary spin flips without a systematic plan, rendering it difficult to derive generalizable insights into the system's physics from their behavior. In contrast, \model~not only reaches the ground state quickly but also employs an elegant and physically interpretable strategy. The algorithm autonomously devises a hierarchical optimization approach by identifying and sequentially flipping the core spins of non-overlapping hexagonal clusters, each composed of six triangular plaquettes. This ``smart" technique satisfies all six plaquettes simultaneously, demonstrating \model’s ability to adopt a global optimization perspective inspired by human-like reasoning in optimizing the $p$-spin glass model. The emergence of such problem-solving behavior, where the algorithm discovers higher-order spin update patterns rather than relying on single-spin flips, provides strong evidence that \model~ develops physically meaningful strategies, surpassing traditional methods in both efficiency and interpretability.

\section*{Conclusion and outlook}
This paper presents \model, a unified machine learning framework that integrates hypergraph neural networks and deep reinforcement learning to efficiently optimize $p$-spin glass models. Through extensive numerical experiments, we demonstrate that \model~consistently outperforms conventional simulated annealing methods, achieving significantly lower energy densities and faster convergence across various lattice structures (triangular, square, and hexagonal) and interaction orders. Additionally, \model~exhibits robust performance under different bond distributions (bimodal vs. Gaussian), varying system scales, and mixed-$p$ interaction models.  

A key strength of \model~lies in its ability to naturally handle high-order interactions without relying on auxiliary spins or quadratization techniques, thus preserving computational efficiency and reducing resource overhead. By incorporating gauge transformations directly into the learning process, \model~ achieves faster training convergence and superior optimization performance compared to baseline methods. Furthermore, \model’s versatility extends beyond spin glass models, achieving state-of-the-art results on NP-hard problems such as random $p$-XORSAT, hypergraph MaxCut, and traditional MaxCut.

The integration of gauge invariance in \model’s architecture and its application to high-order optimization problems open new avenues for physics-guided AI research. Our framework not only advances the theoretical understanding of disordered systems but also provides a powerful tool for real-world combinatorial optimization challenges.

\bibliographystyle{apsrev4-2}
\bibliography{ref}

@article{takada2024ising,
  title={Ising model formulation for highly accurate topological color codes decoding},
  author={Takada, Yugo and Takeuchi, Yusaku and Fujii, Keisuke},
  journal={Physical Review Research},
  volume={6},
  number={1},
  pages={013092},
  year={2024},
  publisher={APS}
}

@article{newman_ground_2021,
	title = {Ground {State} {Stability} and the {Nature} of the {Spin} {Glass} {Phase}},
	url = {http://arxiv.org/abs/2110.11229},
	urldate = {2021-11-16},
	journal = {arXiv:2110.11229 [cond-mat, physics:math-ph]},
	author = {Newman, C. M. and Stein, D. L.},
	month = oct,
	year = {2021},
	note = {arXiv: 2110.11229},
}

@article{charbonneau_fractal_2014,
	title = {Fractal free energy landscapes in structural glasses},
	volume = {5},
	copyright = {2014 Springer Nature Limited},
	issn = {2041-1723},
	url = {https://www.nature.com/articles/ncomms4725},
	doi = {10.1038/ncomms4725},
	language = {en},
	number = {1},
	urldate = {2024-10-18},
	journal = {Nature Communications},
	author = {Charbonneau, Patrick and Kurchan, Jorge and Parisi, Giorgio and Urbani, Pierfrancesco and Zamponi, Francesco},
	month = apr,
	year = {2014},
	note = {Publisher: Nature Publishing Group},
	pages = {3725},
}

@article{gardner_spin_1985,
	title = {Spin glasses with \textit{p}-spin interactions},
	volume = {257},
	issn = {0550-3213},
	url = {https://www.sciencedirect.com/science/article/pii/0550321385903748},
	doi = {10.1016/0550-3213(85)90374-8},
	urldate = {2024-10-18},
	journal = {Nuclear Physics B},
	author = {Gardner, E.},
	month = jan,
	year = {1985},
	pages = {747--765},
}

@article{thomas_optimizing_2011,
	title = {Optimizing glassy p-spin models},
	volume = {83},
	issn = {1539-3755, 1550-2376},
	url = {http://arxiv.org/abs/1010.2524},
	doi = {10.1103/PhysRevE.83.046709},
	number = {4},
	urldate = {2024-01-25},
	journal = {Physical Review E},
	author = {Thomas, Creighton K. and Katzgraber, Helmut G.},
	month = apr,
	year = {2011},
	note = {arXiv:1010.2524 [cond-mat]},
	pages = {046709},
}

@article{gross_simplest_1984,
	title = {The simplest spin glass},
	volume = {240},
	issn = {0550-3213},
	url = {https://www.sciencedirect.com/science/article/pii/0550321384902372},
	doi = {10.1016/0550-3213(84)90237-2},
	number = {4},
	urldate = {2023-12-16},
	journal = {Nuclear Physics B},
	author = {Gross, D. J. and Mezard, M.},
	month = nov,
	year = {1984},
	pages = {431--452},
}

@article{parisi_numerical_2012,
	title = {A numerical study of the overlap probability distribution and its sample-to-sample fluctuations in a mean-field model},
	volume = {92},
	issn = {1478-6435, 1478-6443},
	url = {http://www.tandfonline.com/doi/abs/10.1080/14786435.2011.634843},
	doi = {10.1080/14786435.2011.634843},
	language = {en},
	number = {1-3},
	urldate = {2023-12-12},
	journal = {Philosophical Magazine},
	author = {Parisi, Giorgio and Ricci-Tersenghi, Federico},
	month = jan,
	year = {2012},
	pages = {341--352},
}

@article{sherrington1975solvable,
  title={Solvable model of a spin-glass},
  author={Sherrington, David and Kirkpatrick, Scott},
  journal={Physical review letters},
  volume={35},
  number={26},
  pages={1792},
  year={1975},
  publisher={APS}
}

@article{gyorgyi2001techniques,
  title={Techniques of replica symmetry breaking and the storage problem of the McCulloch--Pitts neuron},
  author={Gy{\"o}rgyi, G{\'e}za},
  journal={Physics Reports},
  volume={342},
  number={4-5},
  pages={263--392},
  year={2001},
  publisher={Elsevier}
}

@article{ghofraniha2015experimental,
  title={Experimental evidence of replica symmetry breaking in random lasers},
  author={Ghofraniha, N and Viola, Ilenia and Di Maria, F and Barbarella, G and Gigli, Giuseppe and Leuzzi, L and Conti, Claudio},
  journal={Nature communications},
  volume={6},
  number={1},
  pages={6058},
  year={2015},
  publisher={Nature Publishing Group UK London}
}

@article{edwards1975theory,
  title={Theory of spin glasses},
  author={Edwards, Samuel Frederick and Anderson, Phil W},
  journal={Journal of Physics F: Metal Physics},
  volume={5},
  number={5},
  pages={965},
  year={1975},
  publisher={IOP Publishing}
}

@article{pierangeli2017observation,
  title={Observation of replica symmetry breaking in disordered nonlinear wave propagation},
  author={Pierangeli, Davide and Tavani, Andrea and Di Mei, Fabrizio and Agranat, Aharon J and Conti, Claudio and DelRe, Eugenio},
  journal={Nature communications},
  volume={8},
  number={1},
  pages={1501},
  year={2017},
  publisher={Nature Publishing Group UK London}
}

@article{roma_ground_2009,
	title = {The ground state energy of the {Edwards}–{Anderson} spin glass model with a parallel tempering {Monte} {Carlo} algorithm},
	volume = {388},
	issn = {0378-4371},
	url = {https://www.sciencedirect.com/science/article/pii/S0378437109002544},
	doi = {10.1016/j.physa.2009.03.036},
	language = {en},
	number = {14},
	urldate = {2022-12-01},
	journal = {Physica A: Statistical Mechanics and its Applications},
	author = {Romá, F. and Risau-Gusman, S. and Ramirez-Pastor, A. J. and Nieto, F. and Vogel, E. E.},
	month = jul,
	year = {2009},
	pages = {2821--2838},
}

@article{wang_comparing_2015,
	title = {Comparing {Monte} {Carlo} methods for finding ground states of {Ising} spin glasses: {Population} annealing, simulated annealing, and parallel tempering},
	volume = {92},
	issn = {1539-3755, 1550-2376},
	shorttitle = {Comparing {Monte} {Carlo} methods for finding ground states of {Ising} spin glasses},
	url = {https://link.aps.org/doi/10.1103/PhysRevE.92.013303},
	doi = {10.1103/PhysRevE.92.013303},
	language = {en},
	number = {1},
	urldate = {2022-12-01},
	journal = {Physical Review E},
	author = {Wang, Wenlong and Machta, Jonathan and Katzgraber, Helmut G.},
	month = jul,
	year = {2015},
	pages = {013303},
}

@book{mezard1987spin,
  title={Spin glass theory and beyond: An Introduction to the Replica Method and Its Applications},
  author={M{\'e}zard, Marc and Parisi, Giorgio and Virasoro, Miguel Angel},
  volume={9},
  year={1987},
  publisher={World Scientific Publishing Company}
}

@book{charbonneau2023spin,
  title={Spin Glass Theory and Far Beyond: Replica Symmetry Breaking after 40 Years},
  author={Charbonneau, Patrick and Marinari, Enzo and Parisi, Giorgio and Ricci-tersenghi, Federico and Sicuro, Gabriele and Zamponi, Francesco and Mezard, Marc},
  year={2023},
  publisher={World Scientific}
}

@article{barahona1982computational,
  title={On the computational complexity of Ising spin glass models},
  author={Barahona, Francisco},
  journal={Journal of Physics A: Mathematical and General},
  volume={15},
  number={10},
  pages={3241},
  year={1982},
  publisher={IOP Publishing}
}

@article{larson2010numerical,
  title={Numerical studies of a one-dimensional three-spin spin-glass model with long-range interactions},
  author={Larson, Derek and Katzgraber, Helmut G and Moore, MA and Young, AP},
  journal={Physical Review B—Condensed Matter and Materials Physics},
  volume={81},
  number={6},
  pages={064415},
  year={2010},
  publisher={APS}
}

@article{sourlas1989spin,
  title={Spin-glass models as error-correcting codes},
  author={Sourlas, Nicolas},
  journal={Nature},
  volume={339},
  number={6227},
  pages={693--695},
  year={1989},
  publisher={Nature Publishing Group UK London}
}

@article{kirkpatrick1983optimization,
  title={Optimization by simulated annealing},
  author={Kirkpatrick, Scott and Gelatt Jr, C Daniel and Vecchi, Mario P},
  journal={science},
  volume={220},
  number={4598},
  pages={671--680},
  year={1983},
  publisher={American association for the advancement of science}
}

@article{earl2005parallel,
  title={Parallel tempering: Theory, applications, and new perspectives},
  author={Earl, David J and Deem, Michael W},
  journal={Physical Chemistry Chemical Physics},
  volume={7},
  number={23},
  pages={3910--3916},
  year={2005},
  publisher={Royal Society of Chemistry}
}

@article{khalil2017learning,
  title={Learning combinatorial optimization algorithms over graphs},
  author={Khalil, Elias and Dai, Hanjun and Zhang, Yuyu and Dilkina, Bistra and Song, Le},
  journal={Advances in neural information processing systems},
  volume={30},
  year={2017}
}

@article{fan2020finding,
  title={Finding key players in complex networks through deep reinforcement learning},
  author={Fan, Changjun and Zeng, Li and Sun, Yizhou and Liu, Yang-Yu},
  journal={Nature machine intelligence},
  volume={2},
  number={6},
  pages={317--324},
  year={2020},
  publisher={Nature Publishing Group UK London}
}

@article{nandi2024margin,
  title={Margin Propagation based XOR-SAT Solvers for Decoding of LDPC Codes},
  author={Nandi, Ankita and Chakrabartty, Shantanu and Thakur, Chetan Singh},
  journal={IEEE Transactions on Communications},
  year={2024},
  publisher={IEEE}
}

@incollection{gupta2006sat,
  title={SAT-based verification methods and applications in hardware verification},
  author={Gupta, Aarti and Ganai, Malay K and Wang, Chao},
  booktitle={International School on Formal Methods for the Design of Computer, Communication and Software Systems},
  pages={108--143},
  year={2006},
  publisher={Springer}
}

@article{massacci2000logical,
  title={Logical cryptanalysis as a SAT problem},
  author={Massacci, Fabio and Marraro, Laura},
  journal={Journal of Automated Reasoning},
  volume={24},
  pages={165--203},
  year={2000},
  publisher={Springer}
}

@article{veldt2022hypergraph,
  title={Hypergraph cuts with general splitting functions},
  author={Veldt, Nate and Benson, Austin R and Kleinberg, Jon},
  journal={SIAM Review},
  volume={64},
  number={3},
  pages={650--685},
  year={2022},
  publisher={SIAM}
}

@inproceedings{brakensiek2021mysteries,
  title={On the mysteries of MAX NAE-SAT},
  author={Brakensiek, Joshua and Huang, Neng and Potechin, Aaron and Zwick, Uri},
  booktitle={Proceedings of the 2021 ACM-SIAM Symposium on Discrete Algorithms (SODA)},
  pages={484--503},
  year={2021},
  organization={SIAM}
}

@article{kwon2020pomo,
  title={Pomo: Policy optimization with multiple optima for reinforcement learning},
  author={Kwon, Yeong-Dae and Choo, Jinho and Kim, Byoungjip and Yoon, Iljoo and Gwon, Youngjune and Min, Seungjai},
  journal={Advances in Neural Information Processing Systems},
  volume={33},
  pages={21188--21198},
  year={2020}
}

@inproceedings{kogan2015sketching,
  title={Sketching cuts in graphs and hypergraphs},
  author={Kogan, Dmitry and Krauthgamer, Robert},
  booktitle={Proceedings of the 2015 Conference on Innovations in Theoretical Computer Science},
  pages={367--376},
  year={2015}
}

@article{alidaee19940,
  title={0-1 quadratic programming approach for optimum solutions of two scheduling problems},
  author={Alidaee, Bahram and Kochenberger, Gary A and Ahmadian, Ahmad},
  journal={International Journal of Systems Science},
  volume={25},
  number={2},
  pages={401--408},
  year={1994},
  publisher={Taylor \& Francis}
}

@article{barahona1988application,
  title={An application of combinatorial optimization to statistical physics and circuit layout design},
  author={Barahona, Francisco and Gr{\"o}tschel, Martin and J{\"u}nger, Michael and Reinelt, Gerhard},
  journal={Operations Research},
  volume={36},
  number={3},
  pages={493--513},
  year={1988},
  publisher={INFORMS}
}

@inproceedings{arora2010efficient,
  title={An efficient graph cut algorithm for computer vision problems},
  author={Arora, Chetan and Banerjee, Subhashis and Kalra, Prem and Maheshwari, SN},
  booktitle={Computer Vision--ECCV 2010: 11th European Conference on Computer Vision, Heraklion, Crete, Greece, September 5-11, 2010, Proceedings, Part III 11},
  pages={552--565},
  year={2010},
  organization={Springer}
}

@article{choi2000solving,
  title={Solving sparse semidefinite programs using the dual scaling algorithm with an iterative solver},
  author={Choi, Changhui and Ye, Yinyu},
  journal={Manuscript, Department of Management Sciences, University of Iowa, Iowa City, IA},
  volume={52242},
  pages={115},
  year={2000}
}

@article{kochenberger2013solving,
  title={Solving large scale max cut problems via tabu search},
  author={Kochenberger, Gary A and Hao, Jin-Kao and L{\"u}, Zhipeng and Wang, Haibo and Glover, Fred},
  journal={Journal of Heuristics},
  volume={19},
  pages={565--571},
  year={2013},
  publisher={Springer}
}

@misc{ye2003gset,
  author = {Yinyu Ye},
  title = {The {G}set Dataset},
  howpublished = {\url{https://web.stanford.edu/~yyye/yyye/Gset/}},
  year = {2003},
  note = {Stanford},
  url = {https://web.stanford.edu/~yyye/yyye/Gset/}
}

@article{tonshoff2022one,
  title={One model, any CSP: graph neural networks as fast global search heuristics for constraint satisfaction},
  author={T{\"o}nshoff, Jan and Kisin, Berke and Lindner, Jakob and Grohe, Martin},
  journal={arXiv preprint arXiv:2208.10227},
  year={2022}
}

@article{barabasi1999emergence,
  title={Emergence of scaling in random networks},
  author={Barab{\'a}si, Albert-L{\'a}szl{\'o} and Albert, R{\'e}ka},
  journal={science},
  volume={286},
  number={5439},
  pages={509--512},
  year={1999},
  publisher={American Association for the Advancement of Science}
}

@inproceedings{feng2019hypergraph,
  title={Hypergraph neural networks},
  author={Feng, Yifan and You, Haoxuan and Zhang, Zizhao and Ji, Rongrong and Gao, Yue},
  booktitle={Proceedings of the AAAI conference on artificial intelligence},
  volume={33},
  number={01},
  pages={3558--3565},
  year={2019}
}

@article{yadati2019hypergcn,
  title={Hypergcn: A new method for training graph convolutional networks on hypergraphs},
  author={Yadati, Naganand and Nimishakavi, Madhav and Yadav, Prateek and Nitin, Vikram and Louis, Anand and Talukdar, Partha},
  journal={Advances in neural information processing systems},
  volume={32},
  year={2019}
}

@article{dong2020hnhn,
  title={Hnhn: Hypergraph networks with hyperedge neurons},
  author={Dong, Yihe and Sawin, Will and Bengio, Yoshua},
  journal={arXiv preprint arXiv:2006.12278},
  year={2020}
}

@inproceedings{gilmer2017neural,
  title={Neural message passing for quantum chemistry},
  author={Gilmer, Justin and Schoenholz, Samuel S and Riley, Patrick F and Vinyals, Oriol and Dahl, George E},
  booktitle={International conference on machine learning},
  pages={1263--1272},
  year={2017},
  organization={PMLR}
}

@article{nishimori1983gauge,
  title={Gauge-invariant frustrated Potts spin-glass},
  author={Nishimori, Hidetoshi and Stephen, Michael J},
  journal={Physical Review B},
  volume={27},
  number={9},
  pages={5644},
  year={1983},
  publisher={APS}
}

@article{wegner1971duality,
  title={Duality in generalized Ising models and phase transitions without local order parameters},
  author={Wegner, Franz J},
  journal={Journal of Mathematical Physics},
  volume={12},
  number={10},
  pages={2259--2272},
  year={1971},
  publisher={American Institute of Physics}
}

@inproceedings{mandal2020compressed,
  title={Compressed quadratization of higher order binary optimization problems},
  author={Mandal, Avradip and Roy, Arnab and Upadhyay, Sarvagya and Ushijima-Mwesigwa, Hayato},
  booktitle={Proceedings of the 17th ACM International Conference on Computing Frontiers},
  pages={126--131},
  year={2020}
}

@article{glos2022space,
  title={Space-efficient binary optimization for variational quantum computing},
  author={Glos, Adam and Krawiec, Aleksandra and Zimbor{\'a}s, Zolt{\'a}n},
  journal={npj Quantum Information},
  volume={8},
  number={1},
  pages={39},
  year={2022},
  publisher={Nature Publishing Group UK London}
}

@article{bybee2023efficient,
  title={Efficient optimization with higher-order ising machines},
  author={Bybee, Connor and Kleyko, Denis and Nikonov, Dmitri E and Khosrowshahi, Amir and Olshausen, Bruno A and Sommer, Friedrich T},
  journal={Nature Communications},
  volume={14},
  number={1},
  pages={6033},
  year={2023},
  publisher={Nature Publishing Group UK London}
}

@article{rosenberg1975reduction,
  title={Reduction of bivalent maximization to the quadratic case},
  author={Rosenberg, Ivo G},
  journal={Cahiers du Centre d'etudes de Recherche Operationnelle},
  volume={17},
  pages={71--74},
  year={1975}
}

@article{hartwig1984recursive, 
title={A recursive branch-and-bound algorithm for the exact ground state of Ising spin-glass models}, 
author={Hartwig, A and Daske, F and Kobe, S}, journal={Computer Physics Communications}, 
volume={32}, 
number={2}, 
pages={133--138}, 
year={1984}, 
publisher={Elsevier} 
}

@article{de1995exact,
  title={Exact ground states of Ising spin glasses: New experimental results with a branch-and-cut algorithm},
  author={De Simone, Caterina and Diehl, Martin and J{\"u}nger, Michael and Mutzel, Petra and Reinelt, Gerhard and Rinaldi, Giovanni},
  journal={Journal of Statistical Physics},
  volume={80},
  pages={487--496},
  year={1995},
  publisher={Springer}
}

@article{grest1986cooling,
  title={Cooling-rate dependence for the spin-glass ground-state energy: Implications for optimization by simulated annealing},
  author={Grest, Gary S and Soukoulis, CM and Levin, K},
  journal={Physical review letters},
  volume={56},
  number={11},
  pages={1148},
  year={1986},
  publisher={APS}
}

@article{lucas2014ising,
  title={Ising formulations of many NP problems},
  author={Lucas, Andrew},
  journal={Frontiers in physics},
  volume={2},
  pages={5},
  year={2014},
  publisher={Frontiers Media SA}
}

@article{schuetz2022combinatorial,
  title={Combinatorial optimization with physics-inspired graph neural networks},
  author={Schuetz, Martin JA and Brubaker, J Kyle and Katzgraber, Helmut G},
  journal={Nature Machine Intelligence},
  volume={4},
  number={4},
  pages={367--377},
  year={2022},
  publisher={Nature Publishing Group UK London}
}

@article{heydaribeni2024distributed,
  title={Distributed constrained combinatorial optimization leveraging hypergraph neural networks},
  author={Heydaribeni, Nasimeh and Zhan, Xinrui and Zhang, Ruisi and Eliassi-Rad, Tina and Koushanfar, Farinaz},
  journal={Nature Machine Intelligence},
  volume={6},
  number={6},
  pages={664--672},
  year={2024},
  publisher={Nature Publishing Group UK London}
}

@article{shen2025free,
  title={Free-energy machine for combinatorial optimization},
  author={Shen, Zi-Song and Pan, Feng and Wang, Yao and Men, Yi-Ding and Xu, Wen-Biao and Yung, Man-Hong and Zhang, Pan},
  journal={Nature Computational Science},
  pages={1--11},
  year={2025},
  publisher={Nature Publishing Group US New York}
}

@article{fan2023searching,
  title={Searching for spin glass ground states through deep reinforcement learning},
  author={Fan, Changjun and Shen, Mutian and Nussinov, Zohar and Liu, Zhong and Sun, Yizhou and Liu, Yang-Yu},
  journal={Nature communications},
  volume={14},
  number={1},
  pages={725},
  year={2023},
  publisher={Nature Publishing Group UK London}
}

@article{nishimori1981internal,
  title={Internal energy, specific heat and correlation function of the bond-random Ising model},
  author={Nishimori, Hidetoshi},
  journal={Progress of Theoretical Physics},
  volume={66},
  number={4},
  pages={1169--1181},
  year={1981},
  publisher={Oxford University Press}
}

@article{kirkpatrick1987connections,
  title={Connections between some kinetic and equilibrium theories of the glass transition},
  author={Kirkpatrick, TR and Wolynes, PG},
  journal={Physical Review A},
  volume={35},
  number={7},
  pages={3072},
  year={1987},
  publisher={APS}
}

@article{bovier2001spin,
  title={The spin-glass phase-transition in the Hopfield model with p-spin interactions},
  author={Bovier, Anton and Niederhauser, Beat},
  journal={arXiv preprint cond-mat/0108235},
  year={2001}
}

@article{nikhar2024all,
  title={All-to-all reconfigurability with sparse and higher-order Ising machines},
  author={Nikhar, Srijan and Kannan, Sidharth and Aadit, Navid Anjum and Chowdhury, Shuvro and Camsari, Kerem Y},
  journal={Nature Communications},
  volume={15},
  number={1},
  pages={8977},
  year={2024},
  publisher={Nature Publishing Group UK London}
}

@article{molnar2018continuous,
  title={A continuous-time MaxSAT solver with high analog performance},
  author={Moln{\'a}r, Botond and Moln{\'a}r, Ferenc and Varga, Melinda and Toroczkai, Zolt{\'a}n and Ercsey-Ravasz, M{\'a}ria},
  journal={Nature communications},
  volume={9},
  number={1},
  pages={4864},
  year={2018},
  publisher={Nature Publishing Group UK London}
}

@article{dinur2005hardness,
  title={On the hardness of approximating minimum vertex cover},
  author={Dinur, Irit and Safra, Samuel},
  journal={Annals of mathematics},
  pages={439--485},
  year={2005},
  publisher={JSTOR}
}

@article{li2018combinatorial,
  title={Combinatorial optimization with graph convolutional networks and guided tree search},
  author={Li, Zhuwen and Chen, Qifeng and Koltun, Vladlen},
  journal={Advances in neural information processing systems},
  volume={31},
  year={2018}
}

@article{ozeki1995gauge,
  title={Gauge transformation for dynamical systems of ising spin glasses},
  author={Ozeki, Y},
  journal={Journal of Physics A: Mathematical and General},
  volume={28},
  number={13},
  pages={3645},
  year={1995},
  publisher={IOP Publishing}
}

@article{hagberg2020networkx,
  title={Networkx: Network analysis with python},
  author={Hagberg, Aric and Conway, Drew},
  journal={URL: https://networkx. github. io},
  pages={1--48},
  year={2020}
}

@misc{gurobi,  author = "Gurobi Optimization, LLC",  title = "Gurobi Optimizer Reference Manual",  year = 2021}

@article{baxter1973exact,
  title={Exact solution of an Ising model with three-spin interactions on a triangular lattice},
  author={Baxter, RJ and Wu, FY},
  journal={Physical Review Letters},
  volume={31},
  number={21},
  pages={1294},
  year={1973},
  publisher={APS}
}

@book{baxter2016exactly,
  title={Exactly solved models in statistical mechanics},
  author={Baxter, Rodney J},
  year={2016},
  publisher={Elsevier}
}

@article{arashiro2003short,
  title={Short-time critical dynamics of the Baxter-Wu model},
  author={Arashiro, Everaldo and de Fel{\'\i}cio, JR Drugowich},
  journal={Physical Review E},
  volume={67},
  number={4},
  pages={046123},
  year={2003},
  publisher={APS}
}

@article{bhattacharya2024computing,
  title={Computing high-degree polynomial gradients in memory},
  author={Bhattacharya, Tinish and Hutchinson, George H and Pedretti, Giacomo and Sheng, Xia and Ignowski, Jim and Van Vaerenbergh, Thomas and Beausoleil, Ray and Strachan, John Paul and Strukov, Dmitri B},
  journal={Nature Communications},
  volume={15},
  number={1},
  pages={8211},
  year={2024},
  publisher={Nature Publishing Group UK London}
}

@article{pedretti2025solving,
  title={Solving Boolean satisfiability problems with resistive content addressable memories},
  author={Pedretti, Giacomo and B{\"o}hm, Fabian and Bhattacharya, Tinish and Heittmann, Arne and Zhang, Xiangyi and Hizzani, Mohammad and Hutchinson, George and Kwon, Dongseok and Moon, John and Valiante, Elisabetta and others},
  journal={npj Unconventional Computing},
  volume={2},
  number={1},
  pages={7},
  year={2025},
  publisher={Nature Publishing Group UK London}
}

\end{document}